\documentclass[sn-standardnature]{sn-jnl}% Standard Nature Portfolio Reference Style
\usepackage{amsmath}  %% for the math formulars - murnaghan 
\usepackage{blindtext}  %% for the step list in the simulation protocol 
\jyear{2021}%

\theoremstyle{thmstyleone}%
\theoremstyle{thmstyletwo}%

\theoremstyle{thmstylethree}%

\usepackage{amssymb}

\begin{document}

\title{%
%Automated uncertainty quantification to determine optimum convergence parameters in plane wave density functional theory calculations via tensor decomposition
Automated optimization of convergence parameters in plane wave density functional theory calculations via a tensor decomposition-based uncertainty quantification}

\author*[1,2]{Jan Janssen}
\email{janssen@mpie.de}
\author[3]{Edgar Makarov}
\author[1,4]{Tilmann Hickel}
\author[3]{Alexander V. Shapeev}
\author[1]{J{\"o}rg Neugebauer}

\affil[1]{%
Max-Planck-Institut f{\"u}r Eisenforschung GmbH,
Max-Planck-Str. 1, 40237 D{\"u}sseldorf, Germany
}
\affil[2]{%
Theoretical Division, Los Alamos National Laboratory, 
Bikini Atoll Rd., SM 30, Los Alamos, NM, USA
}
\affil[3]{%
Skolkovo Institute of Science and Technology, 
Skolkovo Innovation Center, Bolshoy Bulvar, 30/1, Moscow 121205 Russia
}
\affil[4]{%
BAM Federal Institute for Materials Research and Testing, Richard-Willstätter-Str. 11, 12489 Berlin, Germany
}

\abstract{
First principles approaches have revolutionized our ability in using computers to predict, explore and design materials. A major advantage commonly associated with these approaches is that they are fully parameter free. However, numerically solving the underlying equations requires to choose
%-- often manually -- 
a set of convergence parameters. 
%:  Setting these parameters too low may sacrifice the predictive power, selecting them too high may waste valuable computational resources. 
With the advent of high-throughput calculations
%, e.g., to explore large chemical configuration spaces or to build accurate databases for machine learning potentials 
it becomes exceedingly important to achieve a truly parameter free approach.
%by replacing the explicit input of convergence parameters by the targeted precision of a quantity of interest. 
Utilizing uncertainty quantification (UQ) and tensor decomposition we derive a numerically highly efficient representation of the statistical and systematic error in the multidimensional space of the convergence parameters. Based on this formalism we implement a fully automated approach   
that requires as input the target accuracy rather than convergence parameters. The performance and robustness of the approach are shown by applying it to a large set of elements crystallizing in a cubic fcc lattice.   

%In the past, few guidelines and manual benchmarks were often sufficient. However, new applications related to machine learning, finite temperature materials properties and phase transitions reliably require a precision in energy on the order of 1\,meV, which goes well beyond previous criteria. To guarantee this level of precision a detailed uncertainty quantification as well as an automated approach to determine the optimum set of convergence parameters that guarantee a predefined target error is needed. In the present study we therefore decompose and analyse the various error contributions. Based on this insight we derive and implement a numerically robust and fully automatized algorithm that determines the optimum convergence parameters for a given target error. As prototype application we consider convergence checks for bulk properties that are routinely utilized to estimate the accuracy and convergence of density functional theory calculations. Specifically, we consider quantities such as the cohesive energy, equilibrium structure parameters, and bulk modulus obtained from the energy-volume surface for cubic bulk crystals and the $k$-point sampling as well as the plane wave energy cutoff as convergence parameters. The performance and robustness of the approach are shown by applying it to a large set of elements crystallizing in a cubic fcc lattice.      
}


\maketitle

%% \linenumbers

%% main text
\section{Introduction}\label{intro}
Density functional theory (DFT) has evolved as work-horse method to routinely compute for essentially all known materials their properties. 
%A consequence of this success is that DFT-based calculations consume huge amounts of supercomputer resources worldwide \cite{CPU_resources}. Being able to reduce the cost for such calculations, without having to sacrifice their accuracy, would open huge opportunities in saving computational resources as well as electricity.  
While DFT is parameter-free in the sense that no materials specific input parameters are needed, it is not free of numerical convergence parameters. Carefully selecting these parameters is critical: Setting them too low may sacrifice the predictive power, selecting them too high may waste valuable computational resources. Since DFT-based calculations consume huge amounts of supercomputer resources worldwide 
%\cite{CPU_resources} Jan: do you have a good reference?
being able to reduce the cost for such calculations, without having to sacrifice their accuracy, would open large opportunities in saving computational resources.
In the past, the selection of these parameters was based on a few guidelines and manual benchmarks. However, new applications related to machine learning or predicting finite temperature materials properties reliably require a precision in energy on the order of 1\,meV, a precision that goes well beyond previous criteria. 
To guarantee this level of precision a detailed uncertainty quantification as well as an automated approach to determine the optimum set of convergence parameters that guarantee a predefined target error is needed.

Prominent examples of DFT convergence parameters are the number of basis functions and the \textbf{k}-point sampling, which reflect the need to approximate an infinite basis set such as e.g. plane waves (PW) or a continuous set of \textbf{k}-points in the Brillouin zone by discretized finite sets that can be represented on the computer. The accuracy of these approximations can often be described by a single scalar parameter such as the energy cutoff $\epsilon$ or the number of \textbf{k}-points $\kappa$. The total energy surface $E_{tot}(\{\vec{ R_I}, Z_I\}\text{;} \kappa, \epsilon, \dots)$ is thus not only a function of atom coordinates ${\vec{R_I}}$ and species $Z_I$, but also of the convergence parameters $\kappa$, $\epsilon$, etc. Since the total energy surface is the key quantity to derive materials properties, any derived quantity $f[E_{tot}]$ thus depends on the choice of the convergence parameters as well. 

Numerically accurate, i.e. converged, results would be obtained when the convergence parameters approach infinity. In practice, this strategy is not feasible since the required computational resources also scale with the convergence parameters. Therefore, from the beginning of DFT calculations a judicious choice of the convergence parameters was mandatory~\cite{chadi1973, monkhorst1976, dacosta1986, francis1990}. To optimally use computational resources the convergence parameters have to be chosen such that (i) the actual error $\Delta f(\epsilon, \kappa, \dots)$ is smaller than the target error $\Delta f_{\rm target}$ of the quantity of interest and (ii) the required computational resources are minimized. Since the required computational resources scale monotonously  with the convergence parameters the latter condition translates in keeping the convergence parameters as small as possible without violating (i). 

Extensive convergence checks have been mainly reported for simple bulk systems by computing the total potential energy surface (PES) as function of the volume per unit cell~\cite{soler2002, beeler2010, kratzer2019, choudhary2019}. Knowing the energy-volume potential energy surface allows for a direct computation of important materials properties such as the equilibrium lattice structure, mechanical response such as the bulk modulus or the cohesive energy~\cite{murnaghan1944, birch1947, vinet1987}. For more complex quantities derived from the PES such as phonon spectra, surface energies, or free energies routine benchmarks such as by how much a target quantity changes when increasing a specific convergence parameter are common. However, systematic convergence checks are rarely reported, e.g. \cite{grabowski2007, duong2015} for phonons or \cite{grabowski2011} for finite temperature free energies.

\begin{figure}
        \centering
        \includegraphics[width=0.65\textwidth,trim=0cm 0cm 0cm 0cm,clip]{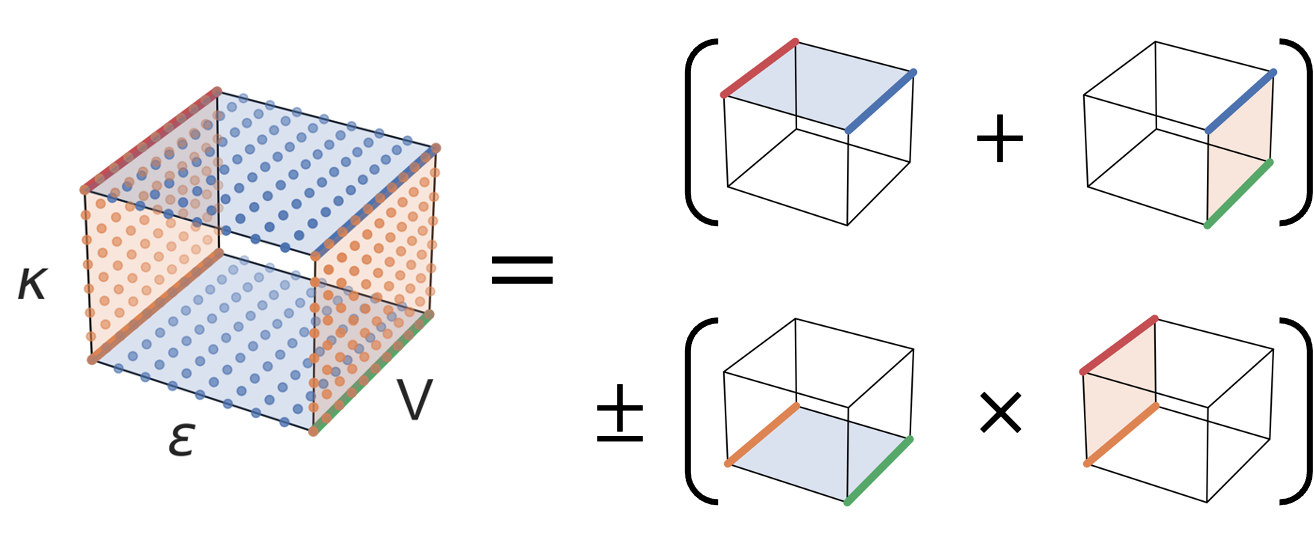}
        \caption{Schematic representation of the discrete mesh of physical and convergence parameters used to map the energy surface $E(V, \epsilon, \kappa)$ (left side) and  graphical representation of the proposed tensor decomposition (right side). The upper/lower part represents the systematic/statistical error contributions given by Eqs.~(\ref{eq:final_energy_decomposition}) and (\ref{eq:factorized_statistical_error}), respectively. To reconstruct the full rank-three tensor on the left only DFT data computed on the orange and blue colored planes is required.}
        \label{fig:motivation}
\end{figure}

In the present study we derive the asymptotic behavior of the systematic and statistical errors considering the energy-volume dependence and derived quantities such as the equilibrium lattice constant and the bulk modulus of cubic materials. We start by computing an extensive set of DFT data $E(V, \epsilon, \kappa)$ spanning the full range of physical (i.e. volume $V$) as well as convergence parameters (i.e., $\epsilon$, $\kappa$). By carefully analysing these data we show in a first step that the rank-three tensor can be decomposed into four rank-two matrices (see Fig.~\ref{fig:motivation}) with full control on systematic and statistical errors. In a second step we develop an efficient tensor decomposition  approach also for derived quantities such as the equilibrium bulk modulus $B_{eq}(\epsilon, \kappa)$. The availability of these extensive and easy-to-compute (via tensor-decomposition) data sets reveals surprising and hitherto not reported correlations between these critical DFT parameters. The derived formalism also allows us to construct and implement a computationally efficient algorithm that in a fully automated fashion predicts optimum convergence parameters that minimize computational effort while simultaneously guaranteeing convergence below a user-given target error.

\section{Construction of an automated approach to determine optimal DFT convergence parameters}

\subsection{Analysis of the DFT convergence parameters}\label{sec:convergence}

As a first step towards an automated uncertainty quantification of the total energy surface and derived quantities we start with an analysis of DFT convergence. As model system we consider the energy-volume curve of bulk fcc-Al constructed by changing the lattice constant $a_{lat}$ of the cubic cell at $T=0$\,K. Due to crystal symmetry all atomic forces exactly vanish, so that the only degree of freedom is the lattice constant or equivalently the volume of the primitive cell ($V=a_{lat}^3/4$). The actual calculations are performed using VASP \cite{kresse1993, kresse1996, kresse19962}. The results and conclusions are not limited to this specific code but can be directly transferred to any plane wave pseudpotential DFT code. 

In this study we focus on the two most relevant convergence parameters of a plane wave (PW) DFT code: the PW energy cutoff $\epsilon$ and the \textbf{k}-point sampling $\kappa$. The integer value $\kappa$ defines the Monkhorst-Pack mesh $(\kappa\times\kappa\times\kappa)$ used to construct the \textbf{k}-point set.

To discuss and derive our approach we first compute and analyze the total energy surface $E(V, \epsilon, \kappa)$
 as function of volume $V$, energy cutoff $\epsilon$ and \textbf{k}-point sampling $\kappa$. We map the energy surface on an equidistant set of $n_V$ volumes $V_i$ ranging over an interval $\pm 10\%$ around the equilibrium volume $V_{\rm eq}$ (see Sec.~\ref{sec:hyper_parameters}), $n_\epsilon$ energy cutoffs $\epsilon_{j}$ and $n_\kappa$ \textbf{k}-point samplings $\kappa_{k}$ (see Fig.~\ref{fig:motivation}). The generated shifted \textbf{k}-point mesh starts at a minimal  \textbf{k}-point sampling of $3\times3\times3$ $(\kappa_{min}=3)$ and goes up to a maximum of $91\times91\times91$ $(\kappa_{max}=91)$ and an energy cutoff of $\epsilon_{min}=200$ eV up to $\epsilon_{max}=1200$ eV. All calculations are executed for primitive cubic cells with a single atom. The \textbf{k}-point sampling is increased in equidistant steps of $\Delta \kappa=2$ to always exclude the Gamma point. The energy cutoff is increased in equidistant steps of $\Delta \epsilon = 20$ eV. For the following analysis we focus on fcc bulk aluminum. Extensive tests for other elements and pseudopotentials show qualitatively the same behavior and will be discussed in Sec.~\ref{benchmark}.
 
\begin{figure}
        \centering
        \includegraphics[width=0.65\textwidth,trim=0cm 0cm 0cm 0cm,clip]{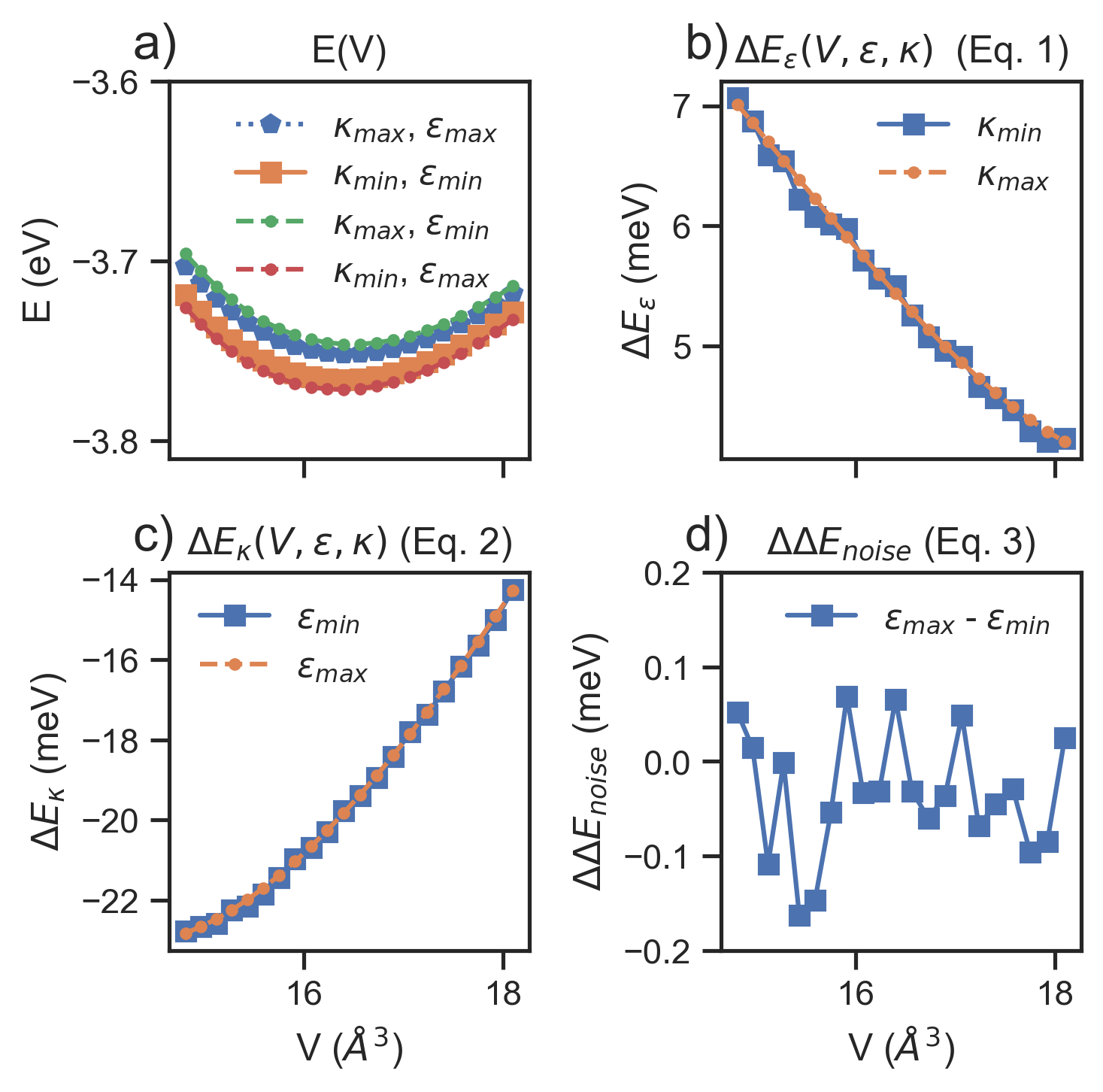}
        \caption{Decomposition of the energy surface $E(V, \epsilon, \kappa)$ into smooth (systematic) and fluctuating (statistic) contributions. (a) Energy-volume curves for a set of minimum and maximum convergence parameters $\epsilon_{min}=260$ eV to $\epsilon_{max}=1200$ eV and $\kappa_{min}=7$ to $\kappa_{max}=91$. (b) and (c)  show the $\epsilon$ and $\kappa$ convergence $\Delta E_\epsilon(V, \epsilon, \kappa)$ and $\Delta E_\kappa(V, \epsilon, \kappa)$ as defined in Eqs.~(\ref{eq:sys_epsilon}) and (\ref{eq:sys_kappa}) for a fixed \textbf{k}-point sampling $\kappa$ and cutoff $\epsilon$, respectively. (d) Residual (statistical) error according to Eq.~(\ref{eq:statistical_error}).}
        \label{fig:energy_differences}
\end{figure}

In total $n_{V}\times n_{\epsilon}\times n_{\kappa}=21\times51\times45=48195$ DFT calculations have been performed for a single pseudopotential. We note already here that such a large number of DFT calculations is not required for the final algorithm. The complete mapping of cutoff and \textbf{k}-point-sampling is used here only to derive and benchmark the asymptotic behavior for $E(V, \epsilon, \kappa)$ and $B_{eq}(\epsilon_i, \kappa_j)\mid_{V_{eq}}$ when going towards large (i.e. extremely well converged) parameters. To analyze the energy surface $E(V, \epsilon, \kappa)$ we define the convergence error with respect to the maximum energy cutoff $\epsilon_{max}$ and \textbf{k}-point sampling $\kappa_{max}$, respectively:
\begin{equation}\label{eq:sys_epsilon}
    \Delta E_\epsilon(V, \epsilon, \kappa) = E(V, \epsilon, \kappa) - E(V, \epsilon_{max}, \kappa)
\end{equation}
\begin{equation}\label{eq:sys_kappa}
    \Delta E_\kappa(V, \epsilon, \kappa) = E(V, \epsilon, \kappa) - E(V, \epsilon, \kappa_{max}) \quad .
\end{equation}
We will first analyze the dependence of the energy over volume of these quantities since this dependence directly impacts convergence of equilibrium quantities such as bulk modulus, lattice constant etc. The energy volume dependence of the energy cutoff convergence $\Delta E_\epsilon$ is shown in Fig.~\ref{fig:energy_differences}(b) for a fixed $\kappa$. As can be seen, using a non-converged energy cutoff $\epsilon_{min}=260$ eV gives rise to a convergence error in the energy that strongly depends on the volume. In the shown example the error is largest for small volumes and monotonously decreases with increasing volume. As a consequence, the equilibrium volume will be shifted to a smaller value compared to the fully converged one. 

Fig.~\ref{fig:energy_differences}(b) also reveals a remarkable and highly useful behavior of the energy cutoff convergence $\Delta E_\epsilon$: It is in first order independent of the \textbf{k}-point sampling. Taking the difference 
\begin{equation}\label{eq:noise}
    \Delta\Delta E_{\rm noise} = \Delta E_{\epsilon}(V, \epsilon, \kappa_2) - \Delta E_{\epsilon}(V, \epsilon, \kappa_1)
\end{equation}
between any two \textbf{k}-point samplings $\kappa_1$ and $\kappa_2$ for a fixed energy cutoff $\epsilon$ results in a volume dependence that resembles random noise. This is shown exemplary in Fig.~\ref{fig:energy_differences}(d) when computing $\Delta\Delta E$ setting $\kappa_1$ and $\kappa_2$ to the minimum and maximum cutoff value: The average $\left< \Delta\Delta E(V) \right>_V$ is approximately zero, the distribution is Gauss-like and any smooth volume dependence is absent. Due to these characteristics we can therefore regard the variance of this contribution 
\begin{equation}
    \label{eq:statistical_error}
    \Delta\Delta E(\epsilon, \kappa) = \sqrt{\left< (\Delta\Delta E_{\rm noise}(V))^2 \right>_V}
\end{equation}
as a statistical error. Its origin are discretization errors arising from discontinuous (discrete) jumps whenever a continuous change in $\kappa$ or $\epsilon$ results in a discontinuous change in the integer number of \textbf{k}-vectors or plane waves.  

Fig.~\ref{fig:energy_differences}(c) shows for the \textbf{k}-point convergence $\Delta E_{\kappa}(V, \epsilon, \kappa)$ an analogous behavior: It is in first order independent of $\epsilon$. The difference $\Delta\Delta E$, is shown in Fig.~\ref{fig:energy_differences}(d) and by construction (see Appendix~\ref{app:stat_error}) identical to the one obtained from the cutoff convergence $\Delta E_{\epsilon}(V, \epsilon, \kappa)$. We can therefore conclude that the energy cutoff and \text{k}-point convergence can be separately considered, i.e., 
\begin{equation}
    \Delta E_\epsilon(V, \epsilon, \kappa) \approx \Delta E_\epsilon(V, \epsilon, \kappa_{max}) \pm \Delta\Delta E(\epsilon, \kappa)
\end{equation}
and
\begin{equation}
    \Delta E_\kappa(V, \epsilon, \kappa) \approx \Delta E_\kappa(V, \epsilon_{max}, \kappa) \pm \Delta\Delta E(\epsilon, \kappa)  .
\end{equation}
The above equations can be summarized in the following expression of the total energy surface:
\begin{equation}\label{eq:final_energy_decomposition}
\begin{split}
     E(V, \epsilon, \kappa) & \approx E(V, \epsilon_{max}, \kappa_{max}) \\
     & + \Delta E_{\epsilon}(V, \epsilon,\kappa_{max}) + \Delta E_{\kappa}(V, \epsilon_{max}, \kappa) \\
     & \pm \Delta\Delta E(\epsilon, \kappa)  .
\end{split}
\end{equation}

The above equation decomposes the original rank-three tensor, which would require to perform $n_V \times n_\epsilon \times n_\kappa$ DFT calculations into three rank-two tensors. The first two require only $n_V \times (n_\epsilon + n_\kappa)$ DFT calculation. The last contribution, $\Delta\Delta E(\epsilon, \kappa)$ requires for each pair of $\epsilon, \kappa$ also the computation at all volumes, i.e., $n_V \times n_\epsilon \times n_\kappa$ DFT calculations. In the following Section we will analyze the statistical error and derive a computationally efficient approach to compute this error.

\subsection{Analysis of the statistical error}

\begin{figure}
        \centering
        \includegraphics[width=0.68\textwidth,trim=0cm 0cm 0cm 0cm,clip]{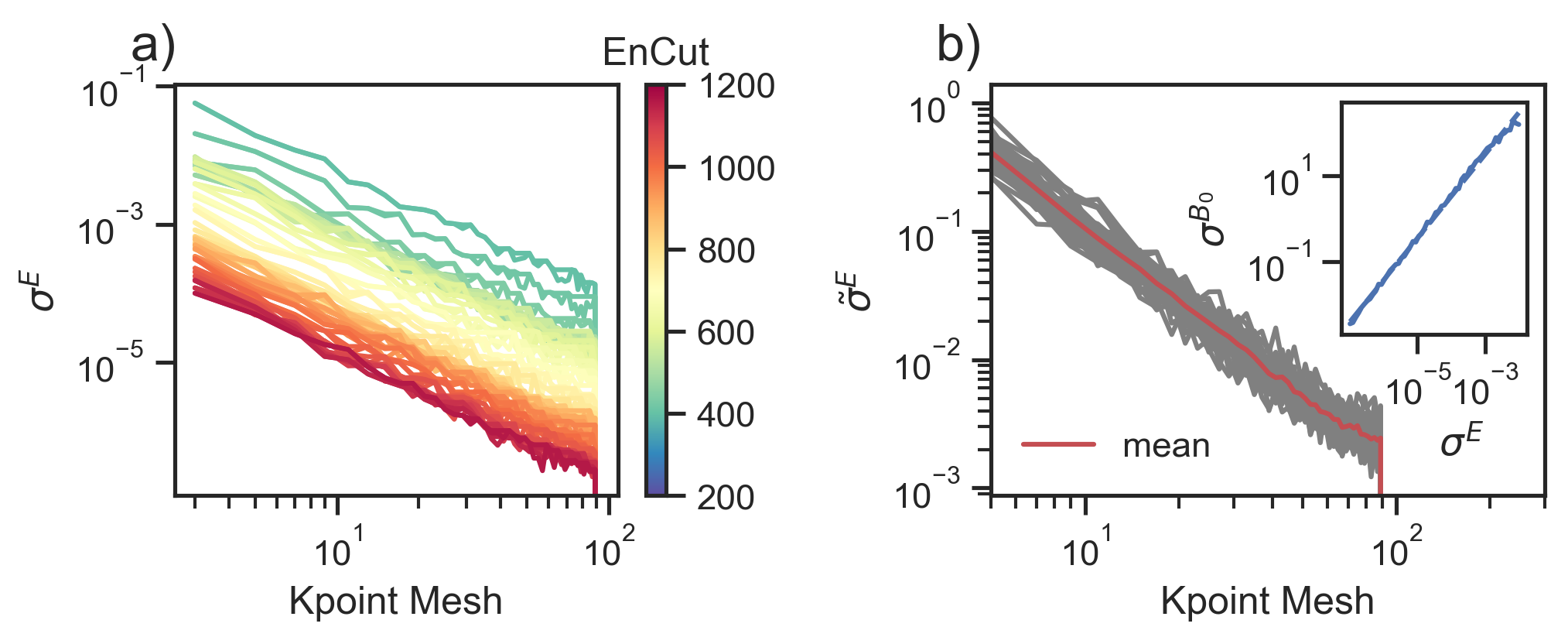}
        \caption{Convergence of the statistical error (Eq.~\ref{eq:statistical_error}) for Cu. a) Convergence over \textbf{k}-point mesh with the different colors denoting the different energy cutoffs. b) Normalized \textbf{k}-point convergence ($\sigma^E(\epsilon, \kappa)/\sigma^E(\epsilon, \kappa_{min})$). The red solid line illustrates the mean. Finally the inset in b) relates the statistical error in energy $\sigma^E$ to an error in the bulk modulus $\sigma^{B_0}$ using bootstrapping (s. Sec.~\ref{sec:decomposition}). }
        \label{fig:statistical_error}
\end{figure}

Having the full dataset of DFT energies as function of V, $\epsilon$ and $\kappa$ allows us to directly compute the statistical error as defined in Eq.~(\ref{eq:statistical_error}). The results are summarized in Fig.~\ref{fig:statistical_error}(a). In the double-logarithmic plot the \textbf{k}-point convergence shows an almost linear dependence, with a vertical shift when changing the energy cutoff (color coded). The fact that the slope remains unchanged when changing the energy cutoff indicates that the \textbf{k}-point convergence of the statistical error is independent of the energy cutoff except for a proportionality factor. To verify this independence we plot the normalized statistical error $\sigma^E(\epsilon, \kappa)/\sigma^E(\epsilon, \kappa_{min})$ in Fig.~\ref{fig:statistical_error}(b). Having this insight we consider the normalized statistical error along the energy cutoff, $\sigma^E(\epsilon, \kappa)/\sigma^E(\epsilon_{min}, \kappa)$. The fact that all curves coincide clearly shows that except for a proportionality factor the cutoff dependence is identical.

Using the above identified empirical relations we can approximate the statistical error by:
\begin{equation}\label{eq:factorized_statistical_error}
    \sigma^E(\epsilon, \kappa) \approx \frac{\sigma^E(\epsilon, \kappa_{min}) \cdot \sigma^E(\epsilon_{min}, \kappa)}{
    \sigma^E(\epsilon_{min}, \kappa_{min})} \quad .
\end{equation}
The computation of $\sigma^E(\epsilon, \kappa_{min})$ and $\sigma^E(\epsilon_{min}, \kappa)$ requires $n_V \times (n_\epsilon + n_\kappa)$ additional DFT calculations. To obtain a sufficiently large magnitude of the statistical error these calculations are performed at the minimum of the convergence parameter set where the statistical noise is largest. 
%This minimum set is equivalent to the VASP recommended energy cutoff and \textbf{k}-sampling. 
As a consequence these extra calculations are computationally inexpensive. To compute the statistical error a second reference next to $\epsilon_{min}$ or $\kappa_{min}$ is needed (see Eq.~\ref{eq:noise}). We find that $\epsilon_{max}$ and $\kappa_{max}$ provide an accurate estimate. These values do not require any additional DFT calculations since they are identical to the ones used to construct the systematic convergence errors $\Delta E_\epsilon$ and $\Delta E_\kappa$ in Eq. (\ref{eq:final_energy_decomposition}). The above formulation allows a highly efficient computation of the statistical error by reducing the computational effort from  $(n_V \times n_\epsilon \times n_\kappa)$ DFT calculations to $2 n_V (n_\epsilon + n_\kappa)$.

\subsection{Analysis and fit of the energy-volume curve}\label{errorcontribution}

Eq. (\ref{eq:final_energy_decomposition}) together with Eq. (\ref{eq:factorized_statistical_error}) provide a powerful and computationally highly efficient approach to compute energy-volume curves $E(V,\epsilon, \kappa)$ for any set of convergence parameters $\epsilon$ and $\kappa$ and are a key result of the present study. They also form the basis for the automated approach to derive optimum convergence parameters that will be derived in the following. The underlying relations and assumption have been carefully validated by computing and analysing an extensive set of pseudopotentials and chemical elements, as presented below. 

\subsubsection{Analysis of the energy-volume curve}

\begin{figure}
        \centering
        \includegraphics[width=0.65\textwidth,trim=0cm 0cm 0cm 0cm,clip]{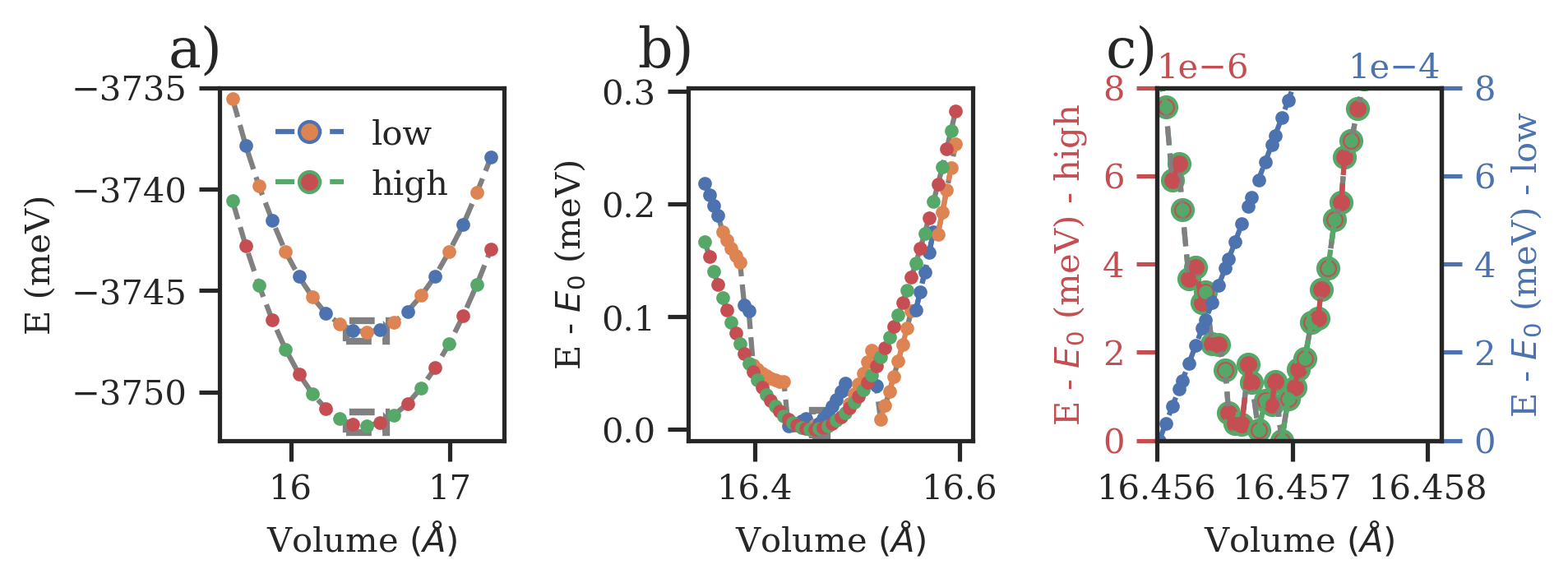}
        \caption{Comparison of two energy-volume curves for fcc-bulk Al. The first one (marked by the blue and orange dots) has been computed using the recommended~\cite{VASP_manual} set of convergence parameters for energy cutoff $\epsilon=240$ eV and \textbf{k}-point sampling $\kappa=11$. The second one (marked by the red and green dots) has been obtained using an extremely high (i.e. well converged) set of parameters ($\epsilon=1000$ eV and $\kappa=101$). Three different volume ranges are shown: (a) $v_{range} =\pm 5\%$ to compare the absolute energies, followed by (b) $v_{range} =\pm 1\%$ to compare the energy changes in reference to the minimum energy and finally (c) $v_{range} =\pm 0.05\%$ again the energy changes. The graphs are coloured based on the number of plane waves, with the color changing whenever the total number of plane waves changes.}
        \label{fig:evcurve}
\end{figure}

Fig.~\ref{fig:evcurve} shows the computed energy-volume curves $E(V, \epsilon, \kappa)$ for two sets of convergence parameters: One with parameters as recommended by the VASP-manual~\cite{VASP_manual} (i.e. $\epsilon_{min}=240$ eV and $\kappa_{min}=11$), the other one for an extremely well converged parameter set. Looking at the results over a large volume range ($\pm 5\%$; Fig.~\ref{fig:evcurve}(a)) the two curves appear to be smooth and well behaved. This may give the impression that the main impact of the convergence parameters is on the absolute energy scale resulting only in a vertical shift. However, going to a $5$ times smaller volume range (Fig.~\ref{fig:evcurve}(b)) the surface with the recommended convergence parameters shows discontinuities that divide the curve. While the segments between two neighboring discontinuities are smooth and analytically well-behaved their boundaries to the neighboring segments are discontinuous in absolute values and derivatives. As a consequence, even a well-defined energy minimum with zero first derivative, which is the definition of the $T=0$~K ground state, does not exist. 

The discontinuous behaviour is a well-known artifact of PW-pseudopotential total energy calculations \cite{francis1990}. The origin is that when changing the volume the number of basis functions (plane waves) changes. Since the number of plane waves is an integer, changing the volume continuously results in discontinuous changes in the number of PWs. Improving the convergence parameters reduces the magnitude of the discontinuity (see line marked by red and green dots in middle figure) but does not remove it (see Fig.~\ref{fig:evcurve}(c) where the volume range has been reduced to $\pm 0.05\%$). Note also that in this interval the lower converged curve (straight line marked by blue dots) has no resemblance at all to the expected close to parabolic energy minimum.

A common strategy to overcome the discontinuous behaviour is to fit the energy-volume points obtained from the DFT-calculations to a smooth fitting function \cite{dacosta1986}. Two main approaches exist: First, using as fit function analytical expressions for the equation of state that describe the relation between the volume of a body and the pressure applied to it \cite{murnaghan1944, birch1947, vinet1987}. From this set we chose the Birch-Murnaghan equation of state, since it is the most popular choice in fitting such energy-volume curves. The second approach is to use polynomial fits. 

In contrast to the discontinuous energy surface, having a smooth fit to the energy-volume data points allows one to obtain the  minimum as well as higher order derivatives around it. Particularly important in this respect are the energy minimum (related to the cohesive energy), the volume at which the energy becomes minimum (equilibrium volume $V_{eq}$ at $T=0$K), as well as the second and third derivative (related to the bulk modulus $B_{eq}$ and its derivative $B^\prime_{eq}$). These quantities can be measured experimentally and thus allow a direct comparison with the theoretical predictions.  

Fitting the data points using either analytical or polynomial functions introduces next to DFT related convergence parameters (e.g. $\epsilon$ and $\kappa$) additional parameters that need to be carefully chosen. For the energy-volume curve these are (i) the number of energy-volume pairs and (ii) the volume range. For a polynomial fit, in addition, also the maximum polynomial degree is a parameter that needs to be tested. Since these parameters are related to the fit and not to the DFT calculation we call them in the following hyper-parameters.  

\subsubsection{Selection of suitable hyper-parameters for the fit}\label{sec:hyper_parameters}

Similarly to the DFT convergence parameters the fit-related hyper-parameters have to be chosen such that (i) the error related to them is smaller than the target error in the physical quantity of interest and (ii) minimize the computational effort (number and computational expense of the necessary DFT calculations). To construct a suitable set of hyper-parameters we consider the bulk modulus:
\begin{align}
    B_{eq} = V \frac{\partial^2 E (V)}{\partial^2 V} \mid_{V=V_{\rm eq}} \quad .
\end{align}
The reason for choosing this parameter is that quantities related to higher derivatives in the total energy surface are more sensitive to fitting/convergence errors. Thus, identifying a set of hyper-parameters for this quantity will guarantee that it works also for less sensitive quantities. Indeed, we checked and validated this hypothesis for quantities related to lower orders in the total energy surface such as equilibrium volume $V_{eq}$ or minimum energy $E_{eq}$.   

We first study the dependence of the bulk modulus with respect to the volume range. As input we use the DFT data set $E(V, \epsilon, \kappa)$ constructed in Sec.~\ref{sec:convergence}, i.e., $21$ energy-volume data points equidistantly distributed over the considered volume range.
For polynomial fits we also tested the impact of the polynomial degree. The accuracy of the fit increases until a maximum degree of $d=11$, i.e., a degree which roughly corresponds to $1/2$ of the number of data points. Going to higher degrees does not reduce the fitting error. 

\begin{figure}
        \centering
        \includegraphics[width=0.6\textwidth,trim=0cm 0cm 0cm 0cm,clip]{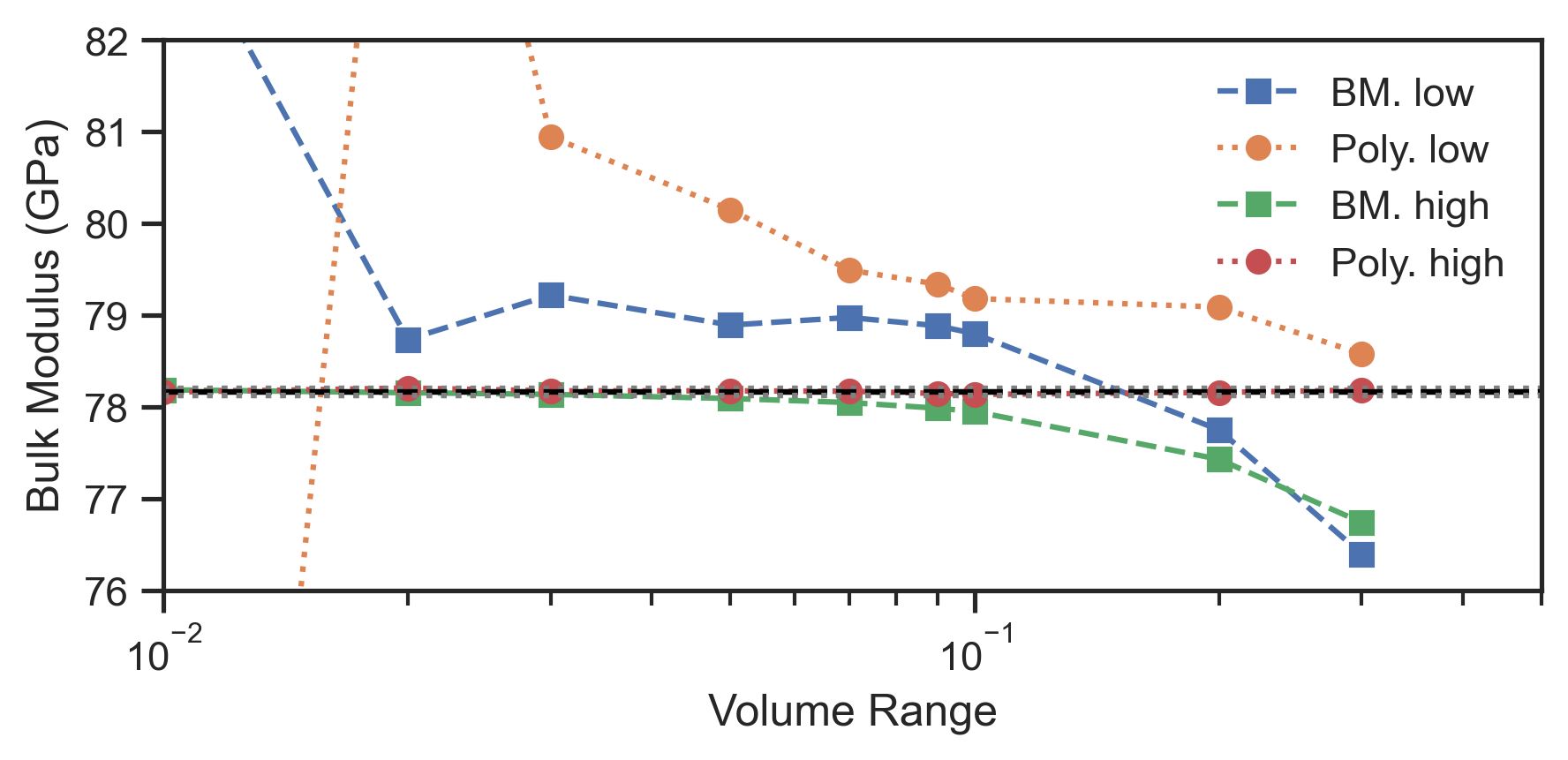}
        \caption{Comparison of the volume range dependence of the bulk modulus for a primitive aluminium supercell with the convergence parameters low ($\epsilon=240$~eV, $\kappa=11$)/high ($\epsilon=1040$~eV and $\kappa=91$) for the Birch Murnaghan (BM.) equation in blue/green in comparison to a polynomial fit in orange/red, each with $N=21$ energy-volume pairs. }
        \label{fig:volumerange}
\end{figure}

The results are summarized in Fig.~\ref{fig:volumerange}. Next to a polynomial fit we also show the results using a physics based analytical expression (Birch-Murnaghan equation of state \cite{birch1947}). As minimum limit for the target error we chose $0.1$\,GPa. We will later show that this target is much smaller than the DFT error related to the exchange-correlation functional, which is on the order of $10$\,GPa and the one related to typical DFT convergence parameters ($\approx 1\dots5$ GPa). 

Fig.~\ref{fig:volumerange} shows that the performance of the two fitting approaches depends on whether low or high convergence parameters are used. For low convergence parameters the analytical fit based on Birch-Murnaghan (blue) is rather insensitive to the exact choice of the volume interval---it remains almost unchanged for intervals between 2 and 10\,\%. Only when going above 10\,\% the underlying analytical model with its four free parameters becomes too unflexible giving rise to an increasing model error. For the commonly recommended interval of $\pm$10\,\% \cite{VASP_manual} the polynomial fit (orange) shows a similar performance but deteriorates both for smaller and larger volume ranges. 

For very high convergence parameters, however, the polynomial approach (red) clearly outperforms the analytical one (green). The polynomial fit is highly robust and largely independent on the chosen volume interval that ranges between $0.1$\% and $30$\%. Also, the number of $21$ energy-volume points is sufficient to achieve the targeted error of $0.1$\,GPa. In this high-convergence regime the analytical Birch-Murnaghan fit provides accurate predictions only for small volume ranges up to ~2\%. 
% We note that common recommendations for the volume range are in the order of $\pm$10\% \cite{VASP_manual}. As shown in Fig.~\ref{fig:volumerange} (dashed blue line), at such large volume ranges a systematic deviation occurs for high convergence parameters. 
The reason is that the analytical expression, which contains only four free fitting parameters is no longer able to adjust to the actual shape of the DFT energy surface. 

Based on this discussion we will use in the following a volume interval of $\pm$10\% and $21$ sample points. With this set of hyper-parameters we verified that the errors arising from the polynomial fit are below the target of $0.1$\,GPA for both the systematic and the statistical error.   
% The error is, however, still well below the common target errors of $\approx 1\dots5\,GPa$ making the choice of a ~$\pm$10\% volume range suitable for lower convergence parameters. 

% ??? Discuss results + impact of the other two parameters (number of points, polynomial degree). Show that our choice (21 points, d=11) gives an error that is well below the minimum error that we consider in the DFT convergence. Give the estimated error for the relevant physical quantities, e.g., $a_{eq}, B_{eq}, B^{\prime}_{eq}$???  

\section{Uncertainty quantification of derived physical quantities}\label{dft}
\subsection{Construction of the error surface}

The compact representation of the energy surface $E(V, \epsilon, \kappa)$ as function of both physical/materials parameters (i.e. volume $V$) and DFT convergence parameters ($\epsilon$, $\kappa$) by Eq.~(\ref{eq:final_energy_decomposition}) together with the set of converged hyper-parameters derived in Sec.~\ref{sec:hyper_parameters} allows us to analyze the convergence of important materials parameters such as equilibrium bulk modulus, lattice constant, cohesive energy etc. with a very modest number of DFT calculations. To test the accuracy and predictive power of the approximate energy surface Eq.~(\ref{eq:final_energy_decomposition}) we first compute the physical quantity from the full set (i.e. $V_i$, $\epsilon_j$, $\kappa_k$) of DFT data. Like in the previous section we focus on the bulk modulus, which is highly sensitive to even small errors. 

The deviation between the bulk modulus and its converged value $B_0(\epsilon_{max}, \kappa_{max})$ as function of the convergence parameters is shown in Fig.~\ref{fig:reconstruction}(a). The color code shows the magnitude of the convergence error in a logarithmic scale. As expected, the error shows a general decrease when going towards higher convergence parameters. The actual dependence, however, is surprisingly complex showing a non-monotonous behavior and several local minima. Performing the same fitting approach on the approximate energy surface Eq.~(\ref{eq:final_energy_decomposition}) (see Fig.~\ref{fig:reconstruction}(c) gives a convergence behavior that shows the same complexity and is virtually indistinguishable from the one shown in Fig.~\ref{fig:reconstruction}(a). Thus, Eq.~(\ref{eq:final_energy_decomposition}) provides a highly accurate and computationally efficient approach for uncertainty quantification of DFT convergence parameters.

It may be tempting to identify the local minima in the error surface as optimum convergence parameters that combine low error with low computational effort. Unfortunately, these local minima are a spurious product of an oscillatory convergence behavior, where at the nodal points the value becomes close to the converged result. Since DFT convergence parameters should be robust against perturbations caused e.g. by changing the shape of the cell, by atomic displacements etc. the local minima are likely to shift, merge or disappear. Thus, selecting parameters based on such local minima would make these parameters suitable only for the exact structure for which the uncertainty quantification has been performed. We therefore construct in the following an envelope function that connects the local maxima. The envelope represents the amplitude of the oscillatory convergence behavior and is roughly independent on the exact position of the nodes (phase shifts).

Fig.~\ref{fig:reconstruction}(b) shows the resulting envelope function. It is much smoother than the original error surface (Fig.~\ref{fig:reconstruction}(a) and \ref{fig:reconstruction} (c)), decreases monotonously when increasing any of the convergence parameters and is free of any spurious local minima. For the further discussion and interpretation of convergence behavior and errors we will exclusively use the envelope function.

\subsection{Error decomposition: Systematic vs statistical contribution}\label{sec:decomposition}

The energy expression given by Eq. (\ref{eq:final_energy_decomposition}) consists of two systematic contributions ($\Delta_\epsilon$ and $\Delta_\kappa$) that smoothly change with the convergence parameters and provide an absolute value.  It also includes a statistical contribution $\Delta\Delta E$, which quantifies the magnitude of the fluctuations around this value. This decomposition into systematic and statistical contributions can be directly transferred to the physical quantities derived from the energy surface. To get the systematic contribution only the systematic part of the energy (i.e. the first three terms in Eq.~(\ref{eq:final_energy_decomposition})) are used as input for the fit. The statistical contribution is obtained using a Monte Carlo bootstrapping approach: The last term in Eq.~(\ref{eq:final_energy_decomposition}) is replaced by a normal distribution $N(\mu, \sigma^2=\Delta\Delta E)$ and the fitting is performed over a large number of such distributions on the best converged surface $E(V, \epsilon_{max}, \kappa_{max}) + N(\mu=0, \sigma^2=\Delta\Delta E(\epsilon, \kappa)$. The inset in Fig.~\ref{fig:statistical_error}(b) shows the computed propagation of the statistical error in total energy $E$ to the statistical error in the bulk modulus.
One can observe a linear relation between the error in the bulk modulus and the magnitude of the noise in the energy-volume curve.
This is because the bulk modulus is obtained as the second derivative of the polynomial fitted to the DFT-computed energy-volume curve.
Because both fitting and taking the second derivative are linear operations, the resulting bulk modulus is a linear functional of the energy values.
Hence, the noise in the bulk modulus scales linearly with the simulated noise in the energy-volume curve.

\begin{figure}
        \centering
        \includegraphics[width=0.68\textwidth,trim=0cm 0cm 0cm 0cm,clip]{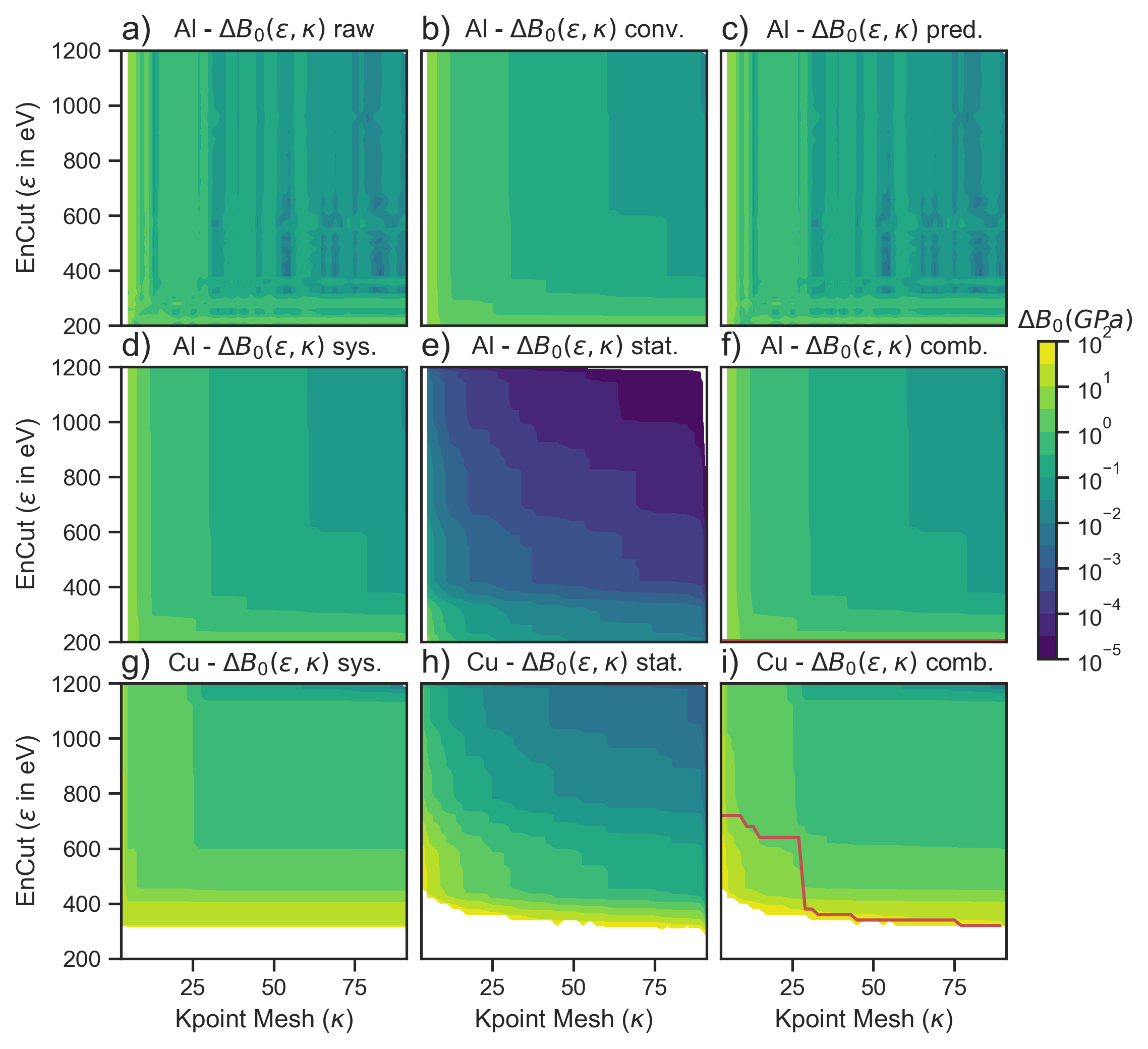}
        \caption{(a-c) Construction and reconstruction of the error surface: a) calculated (raw) error, b) applying the convex hull construction described in the text and c) the reconstruction using Eq.~\ref{eq:final_energy_decomposition}. The middle and bottom row show the convex hull of the systematic error, the statistical error and the total error for Al (d-f) and Cu (g-i). The solid red line in i) is the (phase) boundary separating regions (lower-left part) where the statistical error dominates from regions (upper-right) where the systematic one is large.}
        \label{fig:reconstruction}
\end{figure}

Fig.~\ref{fig:reconstruction} shows the error surface for the statistical error (Fig.~\ref{fig:reconstruction}(a) and (d)), the systematic error (Fig.~\ref{fig:reconstruction}(b) and (e)) as well as the total convergence error (Fig.~\ref{fig:reconstruction}(c) and (f)) for Al and Cu. The two elements have been chosen since they represent the two most different cases that we observe for all investigated potentials.

 \subsection{Error phase diagrams: Interpretation and Application}
 
  Generally, the statistical error becomes dominant for low convergence parameters (in the bottom-left region) while the systematic error dominates for medium and high convergence parameters. The red solid line in Fig.~\ref{fig:reconstruction}(i) shows the boundary where the magnitude of both errors becomes equal. For Al (Fig.~\ref{fig:reconstruction}(c)) this line is absent since the systematic error dominates over the entire region, i.e., even for the lowest considered convergence parameters. Since we have chosen the minimum at or only slightly below the VASP recommended settings the region captures convergence parameters that are commonly chosen. For Cu, in contrast, the statistical error dominates even when using convergence parameters that are well above the recommended ones, i.e., at lower \textbf{k}-point sampling even for an energy cutoff of more than $600$\,eV. 
 
 The boundary (red line in Fig.~\ref{fig:reconstruction}(i)) that separates the regions where either the statistical or the systematic error dominate can be interpreted like a boundary in a phase diagram: It directly provides information about the dominating error (phase) for any given set of convergence parameters (state variables). Such diagrams can be thus regarded as error phase diagram. The knowledge of the error type has direct practical consequences. If the systematic error dominates, the convergence error becomes a simple linear superposition of each individual convergence error (see Eq.~(\ref{eq:final_energy_decomposition})).\footnote{Strictly speaking Eq.~(\ref{eq:final_energy_decomposition}) applies only for the total energy. For derived quantities such as the bulk modulus we observe sizeable deviations. The reason is that next to an explicit dependence of e.g. the bulk modulus on the convergence parameters also an implicit one via the equilibrium volume occurs, i.e. $B_0(\epsilon, \kappa, V_0(\epsilon, \kappa))$}
 It thus allows for any set of ($\kappa$, $\epsilon$) values to determine the deviation from the converged result including its sign. Due to its additive nature the total systematic error will be always dominated by the least converged parameter.
 
 In contrast, the total statistical error can be reduced to any target by just converging a single convergence parameter, which is a direct consequence of its multiplicative rather than additive nature (see Eq.~(\ref{eq:factorized_statistical_error})). It also is the reason why the statistical error decays faster than the systematic error when increasing both convergence parameters simultaneously. 
 
 Knowledge of the above introduced and constructed error phase diagram can be directly used to find convergence parameters that minimize computational resources for a given target accuracy. In the region where the statistical error dominates, the multiplicative nature, i.e., where the targeted accuracy can be achieved by converging only a single parameter, which in practice will be the computationally less expensive one. In typical cases this will be the \textbf{k}-point sampling since the necessary CPU time scales linearly with the number of \textbf{k}-points and the number of \textbf{k}-points decreases with increasing system size. In contrast, if highly converged calculations are desired one will be in the region of the error phase diagram where the systematic error dominates. Since in this region the errors of the individual convergence parameters are additive, converging one parameter better than the other would be a waste of CPU time. Thus, the availability of such error phase diagrams will allow us to provide a highly systematic and intuitive way of identifying optimum sets of DFT parameters. In a way, error phase diagrams may become what thermodynamic phase diagrams are for materials engineers today: Roadmaps for identifying optimum paths (convergence parameters) in materials design (DFT calculations).

 \subsection{Implementation of an automated approach}
 
 Using the concepts outlined in the previous sections allows us to construct an easy to implement automated approach. This approach computes for a given chemical element and its pseudopotential a set of optimum convergence parameters. These optimized parameters guarantee that the error for a user-selected quantity (e.g. bulk modulus, lattice constant etc.) is below a user-defined target error. In the present implementation the developed tool accepts only cubic structures where the unit cell can be fully described by the lattice constant as single variable.
 
 The key steps of the automated approach are as follows:
 \begin{itemize}
     \item Determine the approximate lattice constant at   $\epsilon_{max}$ and $\kappa_{max}$ using the experimental lattice constant, 21 volume points, a volume interval of $\pm 10\%$, and a polynomial fit of order $d=11$.
     \item Perform DFT calculations around the computed equilibrium lattice constant using again $21$ volume points and a volume interval $\pm 10\%$. Compute on each of the volume points the energies along ($\epsilon_{min}$, $\kappa_i$), ($\epsilon_{max}$, $\kappa_i$),  ($\epsilon_i$, $\kappa_{min}$) and ($\epsilon_i$, $\kappa_{max}$).
     \item Use Eq.~(\ref{eq:final_energy_decomposition}) to compute the total energy $E$ on the full 3d mesh (V,$\epsilon$,$\kappa$). No extra DFT calculations are needed.
     \item Compute the systematic and the statistical error of the target quantity $A$ (e.g. bulk modulus) following the discussion given in Sec.~\ref{errorcontribution}. Construct the envelope function of the combined error.
     \item Determine on the error surface $A(\epsilon, \kappa)$ the line $\Delta A(\epsilon_{opt}, \kappa_{opt})=\Delta A_{target}$, i.e., sets of convergence parameters where the predicted error equals the target error. 
     \item Take the set where the curvature of the line is maximum (i.e., close to a 90$^{\rm o}$ angle.
 \end{itemize} 
 
The above algorithm has been implemented in the pyiron framework \cite{janssen2019}. The pyiron-based module requires as input only the pseudopotential and the selection of the target quantity and error. The setup of the DFT jobs, submission on a compute cluster, analysis etc. is done fully automatically without any user intervention.

\begin{figure}
        \centering
        \includegraphics[width=0.68\textwidth,trim=0cm 0cm 0cm 0cm,clip]{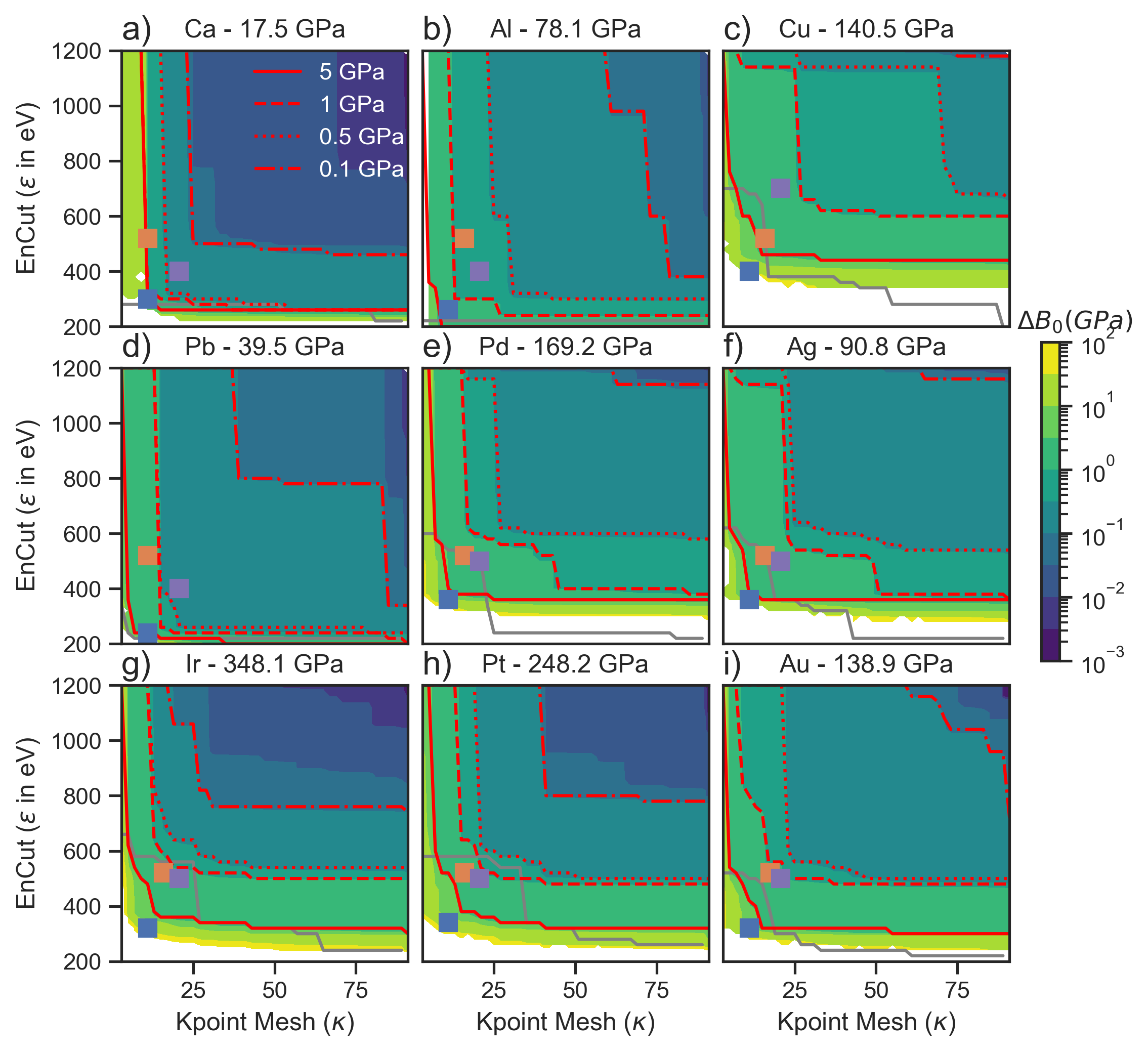}
        \caption{Comparison of convergence for nine fcc metals. The red lines mark iso-contours of constant error (see legend in a). The blue squares mark the parameter set recommended by VASP. The orange and magenta filled squares mark values used in the Materials Project \cite{jain2011, jain2013, jain2015} and the delta project \cite{lejaeghere2016}, respectively. Some elements like Ca a), Al b), Pb d), Ir g), Pt h) and Au i) achieve a precision of up to 0.1 GPa. Others like Cu c), Pd e) and Ag f) are limited to 0.5 GPa in the considered parameter space. Above each contour plot the element and its determined ($\epsilon_{max}, \kappa_{max}$) bulk modulus are given. The grey line denotes the boundary where the systematic and statistical error are equal.}
        \label{fig:summary}
\end{figure} %

\subsection{Application and benchmark of the automated tool}\label{benchmark}

To demonstrate the robustness and the accuracy of the proposed approach we have employed our automated tool to compute error surface and optimized convergence parameters for a wide range of chemical elements and pseudopotentials. Fig.~\ref{fig:summary} shows the convergence behavior for $9$ elements using as target quantity the bulk modulus. For each calculation the PBE-GGA pseudo potential recommended by the VASP manual~\cite{VASP_manual} is chosen. These are also the same pseudo potentials used in the Delta project~\cite{lejaeghere2016}. To visualize the convergence contour lines at target errors of $0.1$ to $5$~GPa are shown on the error surface. Some of the elements such as e.g. Ca allow a convergence down to $0.1$\,GPa even for rather modest convergence parameters. Others, such as e.g. Cu achieve this lower limit barely at the maximum set of convergence parameters studied here. In general, systematic convergence trends that would allow to extract some simple rules for finding optimum convergence parameters are lacking. This emphasizes the importance to provide automatized tools for this task.

For our minimum parameter set the statistical error dominates for all elements except for Al, Ca and Pb as indicated by the grey line in Fig.~\ref{fig:summary}, which marks the boundary where the statistical error and systematic error are equal. Going to errors below 1\,GPa the systematic error becomes the dominating contribution except for elements with a very high bulk modulus such as Ir and Pt. In the region dominated by the systematic error the cutoff and \textbf{k}-point related error contribution are additive (Eq.~(\ref{eq:final_energy_decomposition})). As consequence,  the contour lines in Fig.~\ref{fig:summary} are either parallel to the $\kappa$-axis (i.e. the dominating error is due to the energy cutoff $\epsilon$) or to the $\epsilon$-axis  (with $\kappa$ causing the leading error).

To 'benchmark' the choices made by our automated tool against the choices made by human experts we include the parameters used in two large and well-established high-throughput studies: The Materials Project \cite{jain2011, jain2013, jain2015} (orange squares) and the delta project \cite{lejaeghere2016} (magenta squares). The delta project, which aims at high precision to allow a comparison between different DFT codes, systematically shows an error between $1$ and $5$\,GPa, with a clear tendency towards the $1$\,GPa limit. The Materials Project, where the focus is on computational efficiency and not the highest precision the error is close to the $5$\,GPa limit, for several elements (e.g. Ir, Pt, Au) the error becomes $10$\,GPa or larger.

\section{Conclusions}
In summary, by analyzing the dependence of the total energy not only as function of a single convergence parameter, as commonly done, but as function of a physical parameter (in the present study the volume) as well as multiple convergence parameters (energy cutoff and \textbf{k}-point sampling) simultaneously we identified powerful relations to compute both the statistical and systematic error of the total energy and derived quantities. The identified relations summarized in Eqs.~(\ref{eq:final_energy_decomposition}) and (\ref{eq:factorized_statistical_error}) provide an accurate and computationally efficient tensor decomposition (Fig.~\ref{fig:motivation}), thus allowing to describe error surfaces of multiple convergence parameters for important materials quantities such as e.g. the bulk modulus. 

Being able to construct such energy surfaces for both the systematic and statistical error with a modest number of simple DFT calculations opened the way to construct error phase diagrams which tell whether for any given parameter set one or the other error type dominates. They also provide direct insight of how multiple convergence parameters together affect the errors and allowed us to construct contour lines of constant error. Having this detailed insight is helpful for DFT practitioners to chose and validate accurate yet computationally efficient convergence parameters. It also allowed us to develop and implement a robust and computationally efficient algorithm. The resulting fully automated tool, implemented in pyiron \cite{janssen2019}, predicts an optimum set of convergence parameters with minimum user input, i.e. choice of pseudopotential, desired target error and quantity of interest (e.g. bulk modulus). We expect that this approach where explicit convergence parameters are replaced by a user-selected target error will be particularly important for applications where for a large number of DFT calculations a systematically high accuracy is crucial, e.g. for high-throughput studies and for constructing data sets to be used in machine learning.    

\section{Acknowledgement}
JJ and JN thank Kurt Lejaeghere and Christoph Freysoldt for stimulating discussions and the  
Deutsche Forschungsgemeinschaft (DFG, German Research Foundation) under Projektnummer 405621217 for financial support. 
EM and AVS acknowledge the financial support from the Russian Science Foundation (grant number 18-13-00479).

\section{Competing Interests}
The authors declare no competing interests.

\section{Code Availability}
The Jupyter Notebooks developed to run the calculations and to analyze the data will be provided on our freely accessible pyiron repository (\url{https://github.com/pyiron/pyiron-dft-uncertainty}). The fully interactive Jupyter Notebooks together with our pyiron framework contain the entire code and allow to easily reproduce all calculations and the analysis.

\section{Data Availability}
In addition to the Jupyter Notebooks also the corresponding data will be provided on our freely accessible pyiron repository (\url{https://github.com/pyiron/pyiron-dft-uncertainty}).

\section{Author Contributions}
JJ and JN developed the concepts of the automated uncertainty quantification. JJ implemented the method in the pyiron framework. JJ and EM compared the predictions with existing results. All five authors contributed to the generalization of the method and the writing of the manuscript.

\begin{appendices}
\section{Statistical error}
\label{app:stat_error} 
Equality of Eq.~(\ref{eq:noise}) for both the \textbf{k}-point dependence $\Delta E_{\epsilon}(V, \kappa)$ and the energy cut off dependence $\Delta E_{\kappa}(V, \epsilon)$. For a discrete set of \textbf{k}-points $\kappa_1$, $\kappa_2$ and a discrete set of energy cutoffs $\epsilon_1$, $\epsilon_2$ the energy differences are defined as: 
\begin{align*}
    \Delta E_\epsilon(V, \kappa) &= E(V, \epsilon_{2}, \kappa) - E(V, \epsilon_{1}, \kappa) \\ 
    \Delta E_\kappa(V, \epsilon) &= E(V, \epsilon, \kappa_{2}) - E(V, \epsilon, \kappa_{1}) .
\end{align*}
Based on this definition the statistical error can be defined in two ways:  
\begin{align*}
    \Delta\Delta E_{\rm noise} =& \Delta E_{\epsilon}(V, \kappa_{2}) - \Delta E_{\epsilon}(V, \kappa_{1})\\
    =&  \Big(E(V, \epsilon_{2}, \kappa_{2}) - E(V, \epsilon_{1}, \kappa_{2})\Big) \\
    &- \Big(E(V, \epsilon_{2}, \kappa_{1}) - E(V, \epsilon_{1}, \kappa_{1})\Big) \\
    =& \Big(E(V, \epsilon_{2}, \kappa_{2}) - E(V, \epsilon_{2}, \kappa_{1})\Big) \\
    &- \Big(E(V, \epsilon_{1}, \kappa_{2}) - E(V, \epsilon_{1}, \kappa_{1})\Big) \\
    =& \Delta E_{\kappa}(V, \epsilon_{2}) - \Delta E_{\kappa}(V, \epsilon_{1})
\end{align*}
The statistical error is independent of the order of the differences. 
\end{appendices}

% \bibliography{review}

\end{document}